# Beam Instrumentation for the Tevatron Collider[*]

Ronald S. Moore[#], Andreas Jansson, Vladimir Shiltsev

*Fermi National Accelerator Laboratory, PO Box 500, Batavia, IL 60510 USA*

**Abstract.** The Tevatron in Collider Run II (2001-present) is operating with six times more bunches and many times higher beam intensities and luminosities than in Run I (1992-1995). Beam diagnostics were crucial for the machine start-up and the never-ending luminosity upgrade campaign. We present the overall picture of the Tevatron diagnostics development for Run II, outline machine needs for new instrumentation, present several notable examples that led to Tevatron performance improvements, and discuss the lessons for future colliders.

## INTRODUCTION

Fermilab's Tevatron is currently the world's highest energy proton-antiproton collider operating at 980 GeV per beam. In Collider Run II (2001-present) it operates with six times more bunches and many times higher beam intensities and luminosities than in Run I (1992-1995). The evolutions of the initial and delivered luminosities are shown in Fig. 1. Increasing the total beam intensity made operation of the collider more sensitive to various mechanisms of beam loss that can cause quenches of the superconducting (SC) magnets. Beam orbit drifts, vibrations, diffusion, instabilities and beam-beam effects, as well as electromagnetic long-range and head-on interactions of high intensity proton and antiproton beams have been significant sources of beam loss and lifetime limitations [1]. Precise knowledge of various beam parameters is needed to understand how to fight these phenomena. Naturally, the Tevatron Collider luminosity progress has resulted from optimization of machine performance and reduction of beam losses. Fig. 2 a) shows that early in Run II, combined beam losses only in the Tevatron itself (the last accelerator out of total 7 in the accelerator chain) claimed significantly more than half of the integrated luminosity. Thanks to various improvements [1, 2], including beam diagnostics, losses have been reduced significantly and reached some 30-40% in 2005-2006, paving the road to a many-fold increase of the luminosity. Fig. 2 b) shows the typical evolution of beam intensities at the beginning of a standard cycle of the Tevatron in a period of stable operation.

In this paper we present the most important Tevatron beam diagnostics developments that boosted collider performance, summarize the lessons learned and discuss their applicability to the next large colliders.

___________________
[*]Work supported by Fermi Research Alliance, LLC under Contract No. DE-AC02-07CH11359 with the United States Department of Energy.
[#]ronmoore@fnal.gov

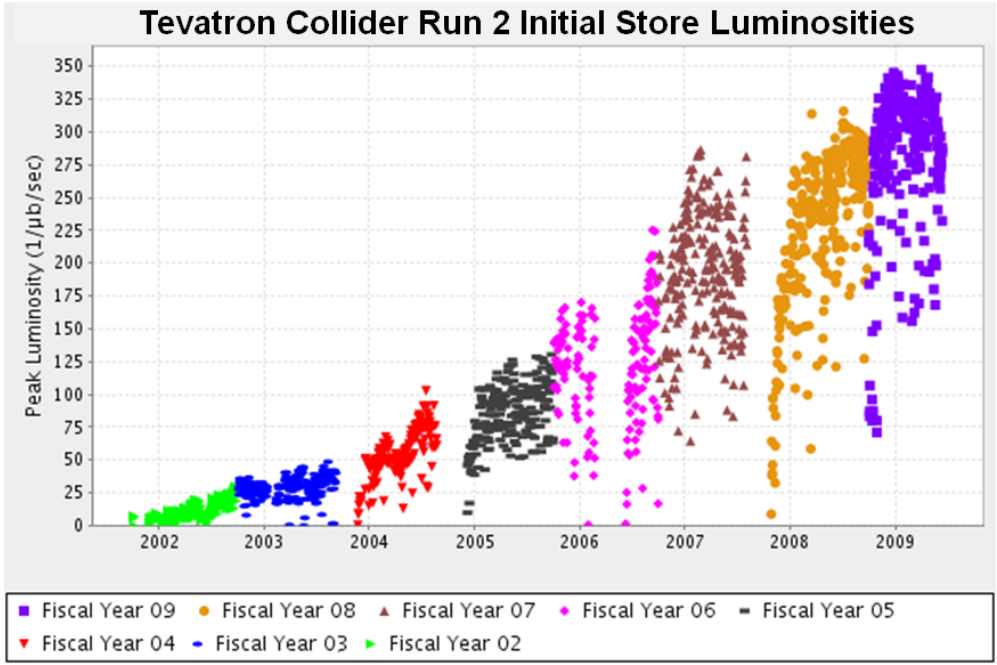

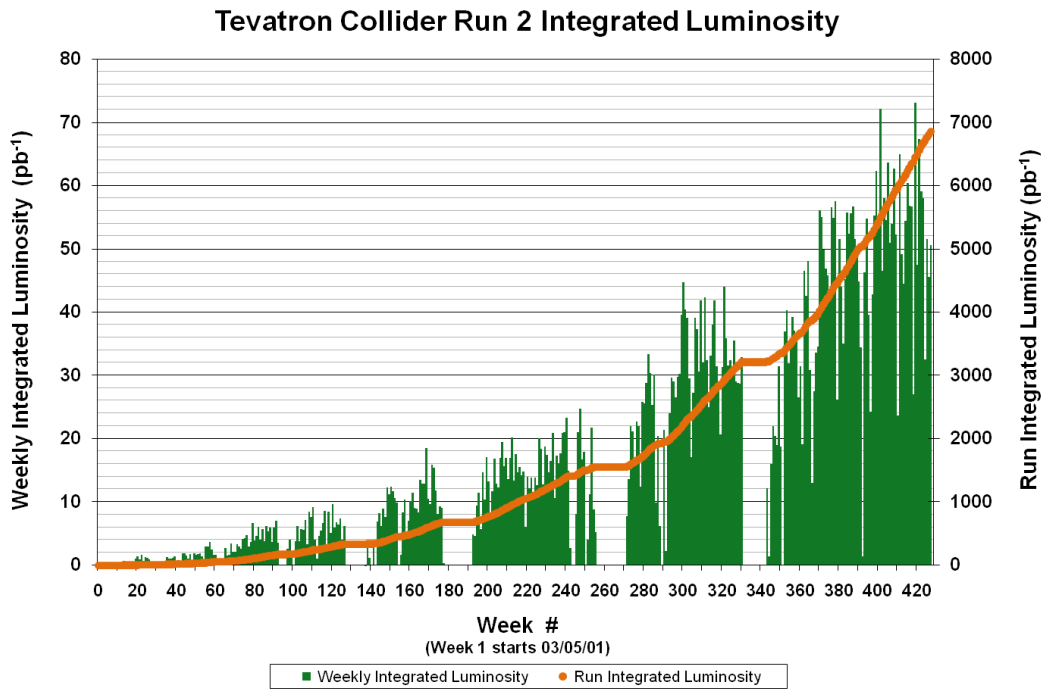

**FIGURE 1.** a) (top) Evolution of initial luminosities of all Tevatron HEP stores since April 2002. b) (bottom) Evolution of weekly and total integrated luminosity delivered to the experiments from March 2001 through June 2009.

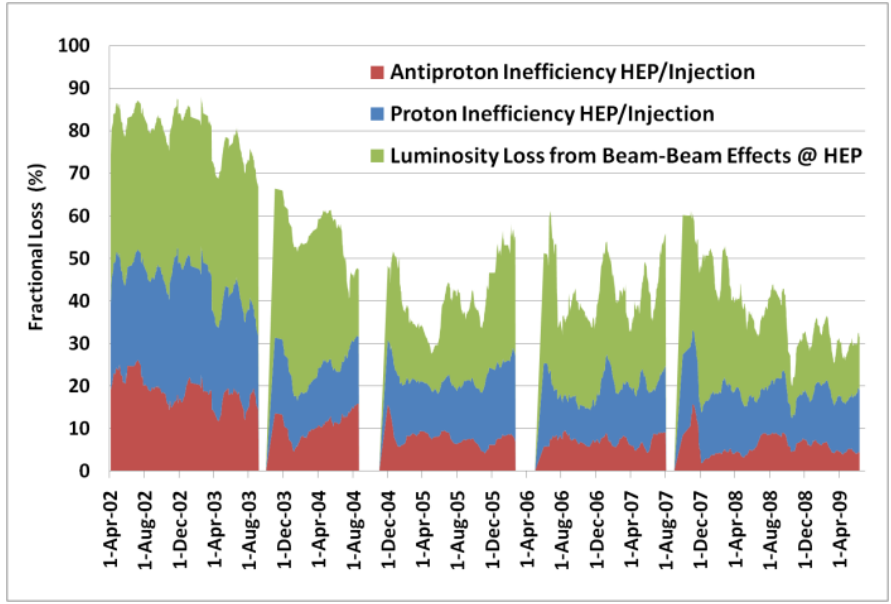

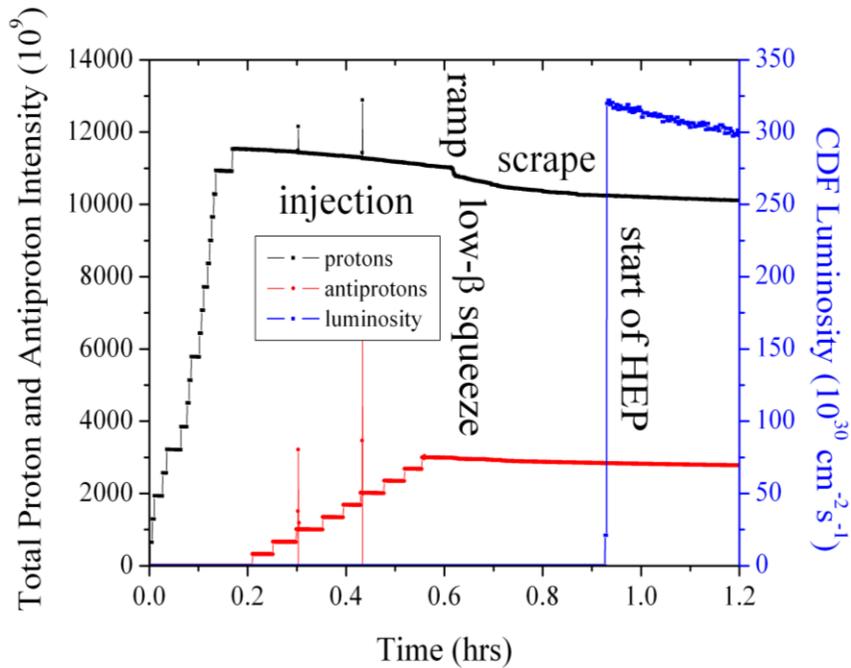

**FIGURE 2.** a) (top) Evolution of beam losses (left axis) in 2002-2009. Red shows fractional loss of antiprotons between injection into the Tevatron and start of collisions, next (blue) one is for loss of protons, green − fractional reduction of the luminosity integral caused by beam-beam effects in collisions. b) (bottom) Injection process and beginning of the luminosity run in store #7040 (May 11, 2009). The square dots are the total proton and antiproton bunch intensities, respectively, as measured by the Fast Bunch Integrator (FBI) system. The line on the right represents the start of the HEP store with an initial peak luminosity of $321\times10^{30}$ cm$^{-2}$ s$^{-1}$. The spikes in the beam intensities are instrumentation artifacts that occur when antiproton bunches pass through proton bunch integration gates during longitudinal cogging [1].

# BEAM DIAGNOSTICS DEVELOPMENTS

Operation of a superconducting magnet hadron collider, like Fermilab's Tevatron, requires a great deal of care, understanding of beam conditions, and trust in the beam diagnostics, because comparatively innocent little imperfections can lead to either beam blow-up and luminosity loss or to beam loss and quench of SC magnets. In the Tevatron such a quench results in 2-4 hours of magnet recovery time and up to 8-16 hours of no-luminosity time needed to produce the antiprotons needed for the next High Energy Physics (HEP) store. Over 8 years of operations we witnessed machine downtimes due to 0.5-1% of beam intensity loss, poor beam lifetime, 0.5-1 mm orbit error, collimator malfunctioning, sequencer error, excursions of tunes or coupling of the order of few 0.001 or several units of chromaticity, instability occurrences, or malfunctioning of kickers, high voltage electrostatic separators, or one of hundreds of power supplies, etc. Naturally, these peculiarities were reflected in the kinds of beam diagnostics we developed (e.g. minimization of their invasiveness) and the way they were exploited (fast data-logging, convenience for post-mortem analysis, etc.).

## Upgrade of Beam Position Monitors

In the Tevatron, the protons and antiprotons circulate within a single beam pipe, so electrostatic separators are used to kick the beams onto distinct helical orbits to allow head-on collisions only at the desired interaction points. (See Fig. 3.) At 150 GeV, separation is limited to ~(10-22) mm by physical aperture, while the separation above 600 GeV, ~(3-6 mm), is limited by the breakdown (spark) rate of the separators at high voltage. Long-range beam effects degrade beam lifetime and machine performance, and having a good model of the optics is essential to understanding problems and how to improve operations. Reliable BPMs with good resolution are needed to measure the optics and construct the model.

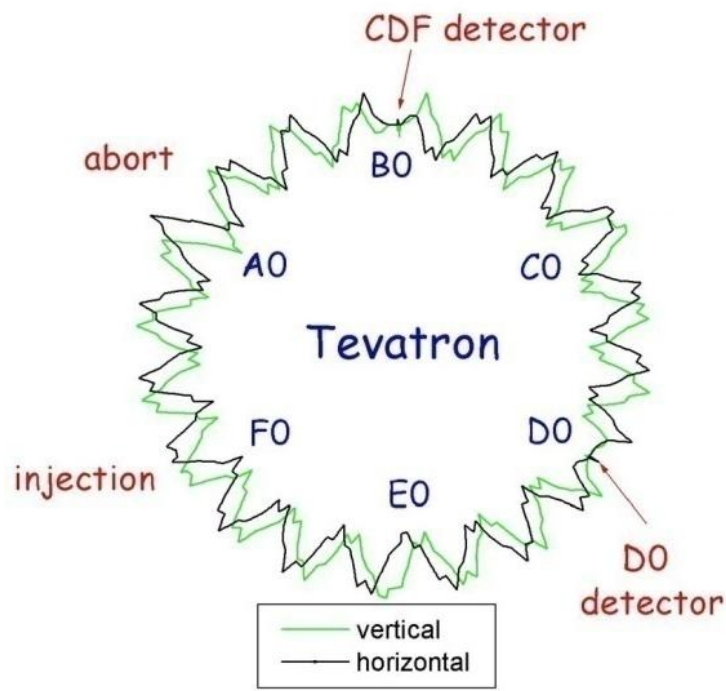

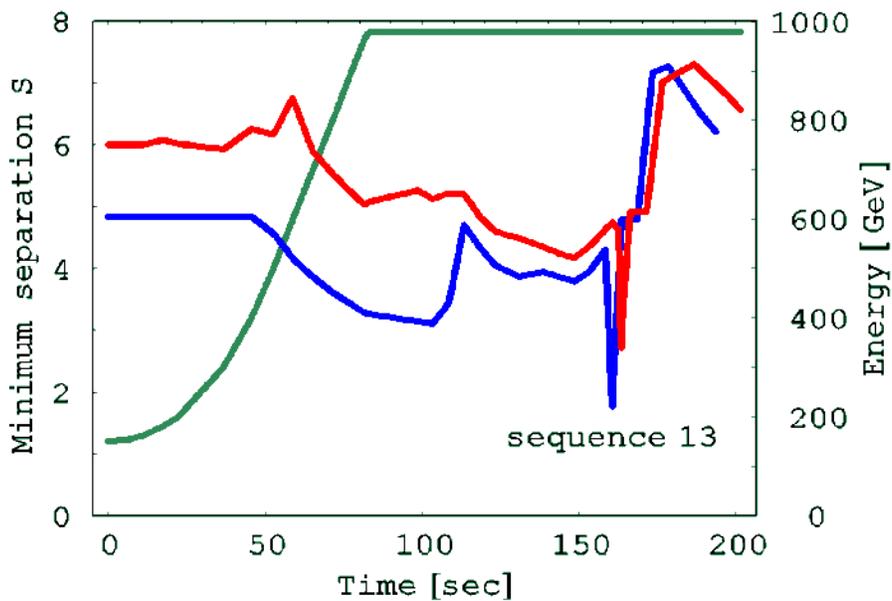

**FIGURE 3.** a) (left) Illustration of the proton helical orbits around the Tevatron. Head-on collisions occur only at the B0 (CDF) and D0 interaction points. b) (right) Comparison of the minimum radial separation between the protons and antiprotons for an old (lower) and present (upper) helix scheme during acceleration and low-beta squeeze. At the point labeled "sequence 13", beam separation reaches a minimum because of a change in the helix orientation needed for HEP. For the old helix, up to 25% of the antiprotons could be lost at that point. Improvements have allowed greater separation at all stages, and the drastic loss of antiprotons was eliminated.

There were several problems with the previous BPM system that hampered machine operations and diagnosis of possible problems. The orbit would deviate significantly from the desired reference orbit, 0.5 mm RMS (root mean square) differences in only 1-2 weeks, so global orbit smoothing was needed regularly. The BPM response to coalesced beam (a transfer concentrated in a single 53 MHz bucket, used for HEP stores) and uncoalesced beam (30 or so consecutive buckets, used for tune-up) differed enough so that a direct comparison between orbits recorded during HEP stores could not be compared easily to proton-only stores used to tune the machine or do orbit smoothing. The BPM position resolution was only ~150 μm and limited optics measurements to at best 20% uncertainty. The turn-by-turn (TBT) capability was unreliable, and the system was blind to antiprotons. All these issues motivated the decision to upgrade the BPM electronics and take advantage of current technology [3, 4]. The 240 Tevatron BPM pickups [5] remained unchanged.

Fig. 4 shows a block diagram of the upgraded BPM electronics system. RG-8 coaxial cables carry the signals from both ends of each pair of BPM pickups to VME racks in service buildings. In the VME crate are analog filter boards (53 MHz bandpass and attenuation), Echotek 8-channel 80 MHz digital receiver boards (ECDR-GC814-FV-A), as well as a Motorola MVME-2400 processor and a module providing clock and trigger signals. The new electronics were installed and commissioned bit by bit, usually between HEP stores, so that only a small number of BPMs would be affected at any one time. This strategy minimized the impact on operations and led to a successful implementation of the new system.

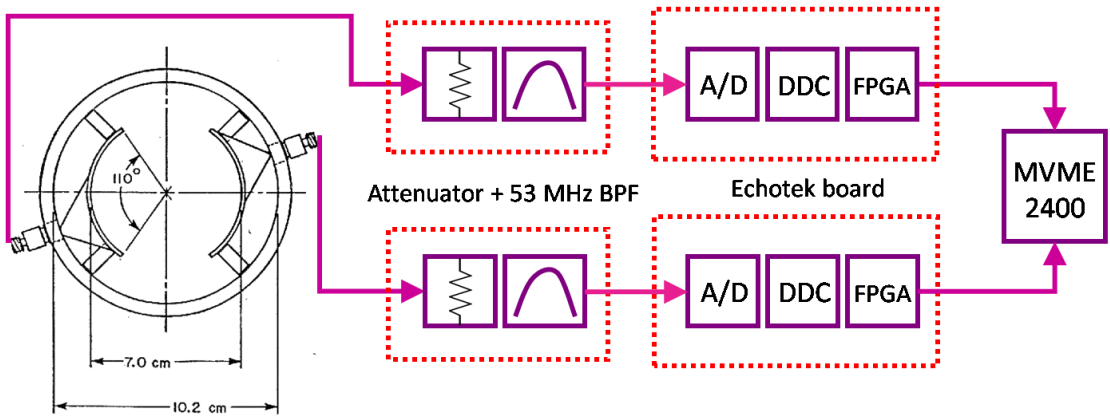

**FIGURE 4.** Block diagram of the upgraded Tevatron BPM electronics. Signals from the BPM pickups go through a 53 MHz band-pass filter (BPF) before being digitized and down-converted on an Echotek model ECDR-GC814-FV-A digital receiver board. A Motorola MVME-2400 processor provides the interface to ACNET for BPM readings and configuration control. Figure adapted from [3].

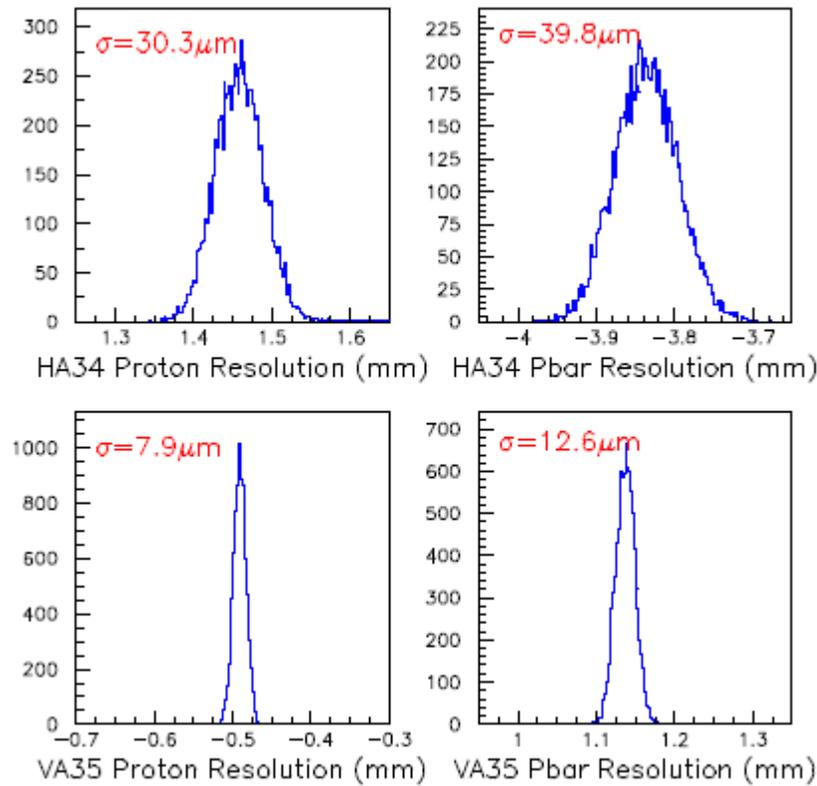

**FIGURE 5.** a) (top) Simultaneous horizontal position measurements from one BPM for protons (left) and antiprotons (right). b) (bottom) Simultaneous vertical position measurements from one BPM for protons (left) and antiprotons (right). The given RMS values include all effects: resolution of the BPM and electronics, real beam motion (especially synchrotron oscillations for the horizontal data), and imperfect cancellation of the proton contamination into the antiproton signal.

An example of the improved resolution of the new BPM electronics is shown in Fig. 5 [6]. The plots show distributions of the proton and antiproton closed orbit position measurements for one horizontal and one vertical BPM. The noted RMS values include the effects of true beam motion, *e.g.* synchrotron oscillations, and the imperfect cancellation of the proton signal onto the signal from the smaller intensity antiprotons. The intrinsic resolution from the BPMs themselves is ≈5 μm, much better than the 150 μm resolution of the old system.

The new electronics provide up to 8192 TBT position measurements at injection and on-demand. Fig. 6 shows TBT measurements from one horizontal and one vertical BPM after intentionally kicking the beam horizontally in order to measure coupling during machine tune-up. The effect from coupling and synchrotron oscillations are clearly visible. The TBT capabilities are being exploited to develop faster and more reliable methods of measuring and correcting the beam optics.

The improved resolution of the BPMs has allowed better measurements of the machine optics which has led to lattice corrections and improvements. For example, the beta functions are now measured to better than 5% accuracy, and a new low-beta lattice with smaller β* was created to increase luminosity by ≈10% [7].

We have observed interesting, rapid beam orbit motion during stores caused by motion of the low-beta quadrupoles and have been able to understand the source and implement an automated orbit-smoothing algorithm [8] that keeps the orbit from wandering during stores (see Fig. 7). In addition, the BPM response no longer depends on the bunch structure, so we can use orbit data from HEP stores to make global orbit corrections when needed.

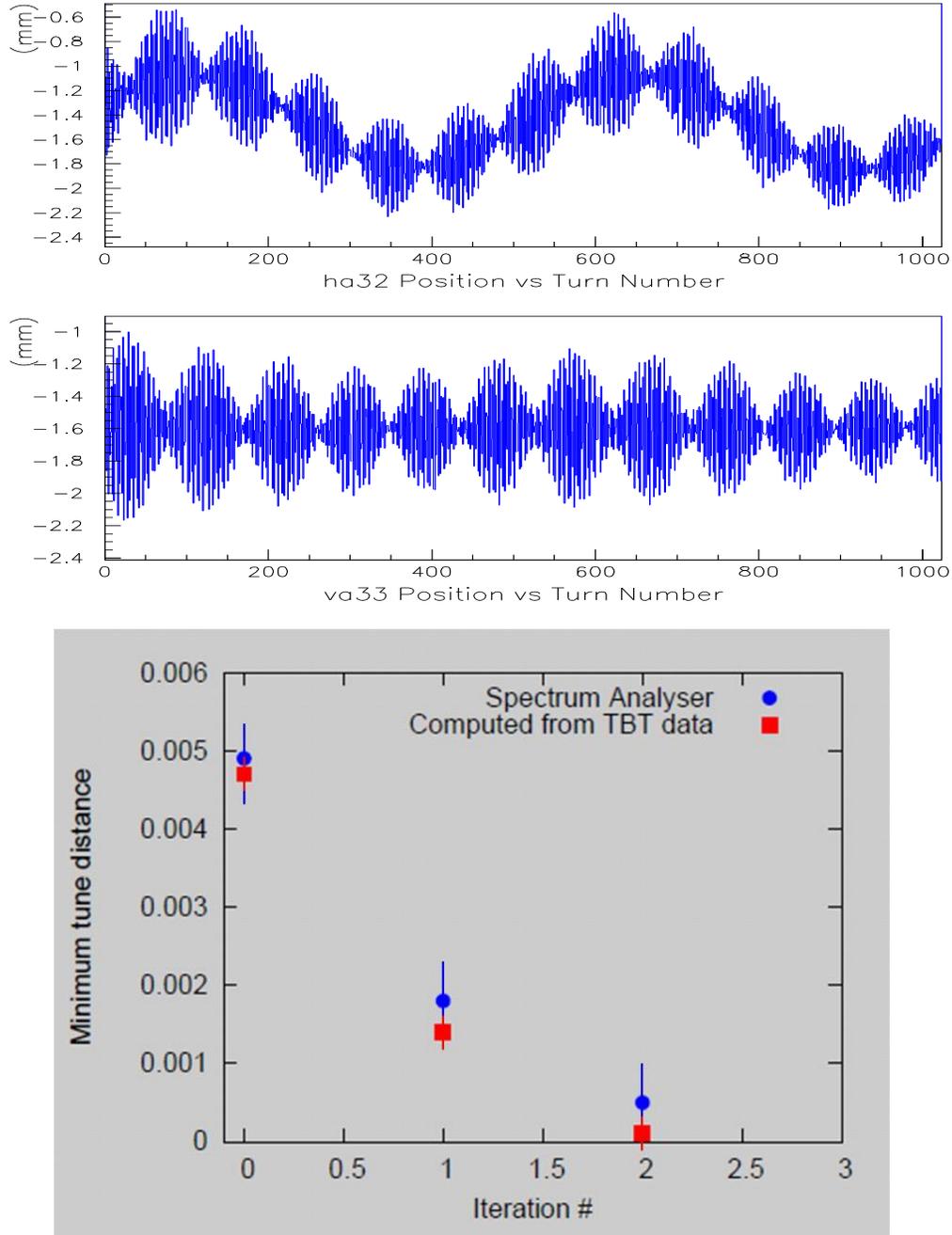

**FIGURE 6.** a) (top) Example of turn-by-turn measurements (top – horizontal, bottom – vertical) from the upgraded BPM electronics after intentionally kicking the proton beam. b) (bottom) Measurements of minimum tune split during attempts to reduce coupling during machine tune-up. The blue points are data from tune measurements made by looking at Schottky signals using a spectrum analyzer, while the red points are derived from turn-by-turn BPM measurements after kicking the beam. The turn-by-turn measurements achieve better results more quickly and more reliably than the spectrum analyzer method.

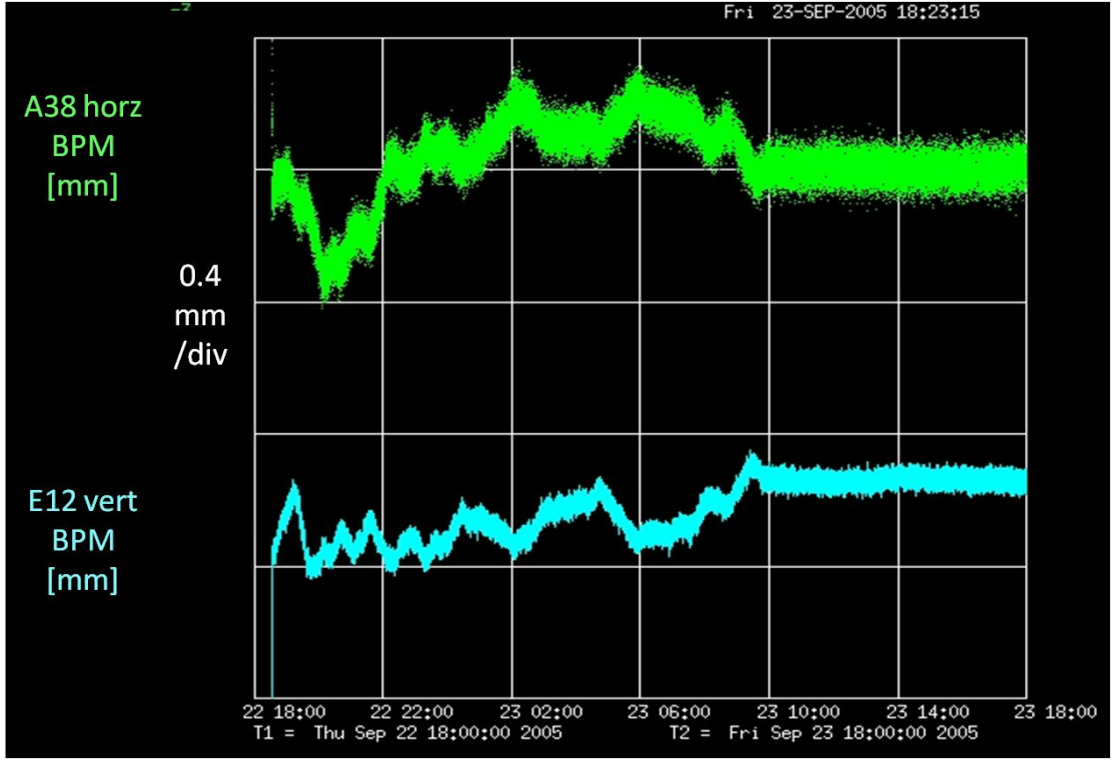

**FIGURE 7.** Proton beam positions from a horizontal BPM (top) and a vertical BPM (bottom) over a 24-hour period during a high-energy physics store. The vertical scale is 400 μm per division. Before an orbit stabilization algorithm was enabled at 10:00, the orbit could wander by over 400 μm in a short period. After orbit stabilization was turned on, the orbit drift was reduced successfully to less than 50 μm. The algorithm uses several dipole correctors near the interaction regions to counteract motion of the low-beta quadrupoles caused by thermal and pressure differences between the Tevatron tunnel and the experimental halls.

## Extra Diagnostics for Low Beta Quadrupoles and IPs

As mentioned above, vibrations of low-beta quadrupoles are primarily responsible for orbit oscillations in the Tevatron, so we equipped each of the magnets with a fast 1 μrad resolution tiltmeter and 0.1 μm resolution hydrostatic level sensors (HLS) to detect vertical motion [9].

Some remarkable examples of orbit and magnet vibrations excited by fire trucks passing by the CDF building and remote earthquakes are shown in Fig. 8 a) and b). The HLS systems also track magnet motion due to continuous sinking of the CDF detector with rate of 0.25-0.5 mm/yr. Such movements lead to a slow drift of the interaction point (IP) position inside the CDF silicon vertex detector (SVX). This and other beam-related information (like loss rates of various counters) can be monitored by Tevatron operators and physicists. For example, both CDF and D0 detectors provide data on the beta-functions at the IPs [10] (see Fig. 9) which is very helpful for us and provides an additional insight into beam collision effects. Vertex analysis also allows separate determination of the proton and antiproton RMS bunch lengths [11].

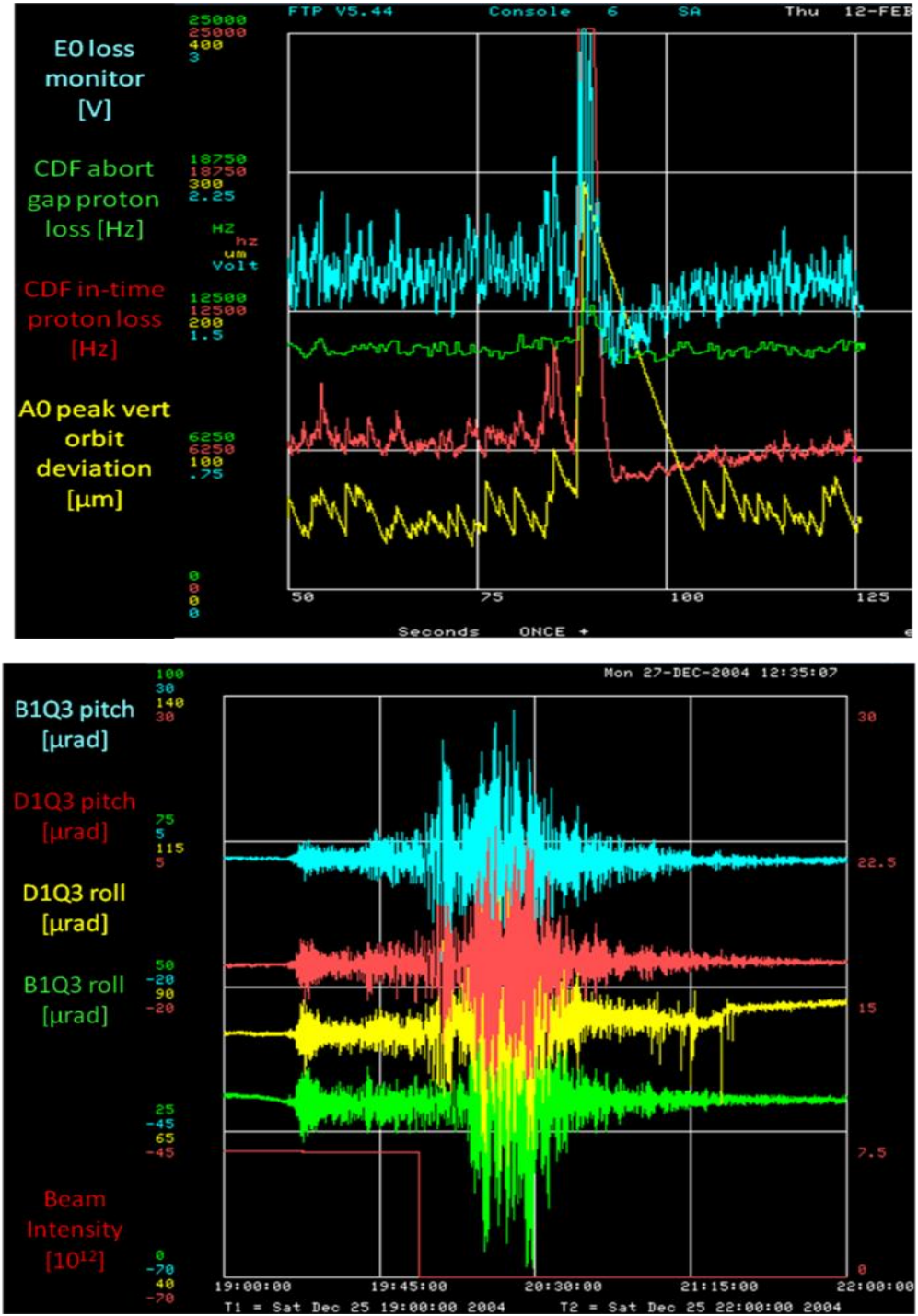

**FIGURE 8.** a) (left) 200 μm orbit oscillations, beam losses and low-beta quadrupole vibrations excited by a 40,000 lb fire truck passing near the CDF Detector Hall. The effect was greatly reduced after installation of new quad girder supports in the FY05 shutdown; b) (right) Disastrous M8.9 earthquake

in Sumatra Dec 25, 2004 resulted in ±50 μrad motion as seen by the tiltmeters on CDF and D0 low-beta quadrupoles. The Tevatron beam (lower red line with a step down) was intentionally terminated before the arrival of the S-wave. The event lasted over 2 hours.

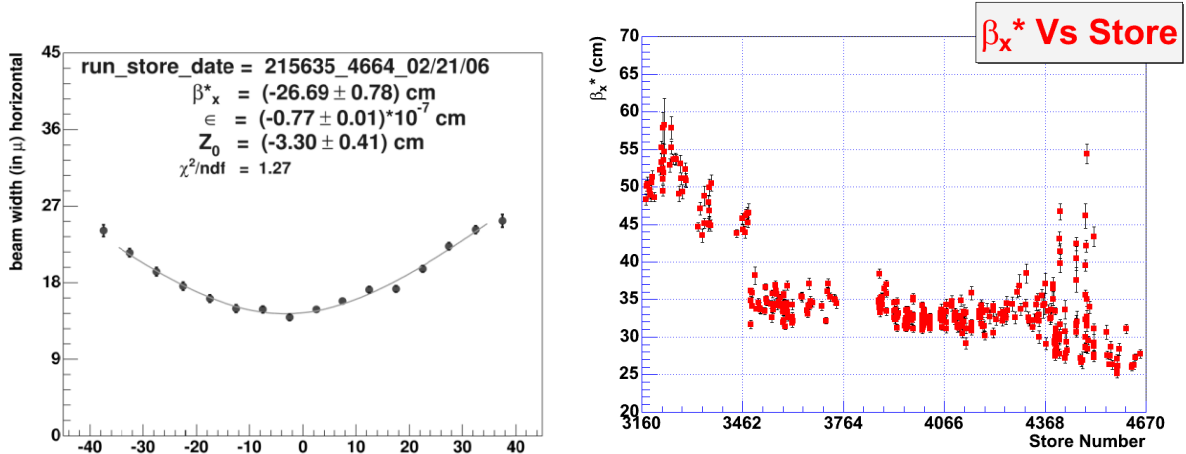

**FIGURE 9.** a) (left) RMS horizontal width of D0 luminous region vs longitudinal position. The parabolic fit is for hour-glass effect with beta*_x=27 cm; b) (right) Jan'04-Mar'06 history of the horizontal beta-function at the D0 IP measured by the D0 silicon vertex detector.

## Tune Diagnostics

There are several systems that measure tunes in the Tevatron. The 21 MHz Schottky is the workhorse for tune measurements during shot setup and studies. The tune is determined by the operator, looking at the Schottky spectrum on a signal analyzer in the Control Room – see Fig. 10. The result may be somewhat subjective, since the spectra typically contain numerous coherent peaks, and it may not be immediately clear which one (if any) represents the real tune. To enhance the signal and make the tunes visible, noise can be injected into the beam through the transverse damper system. The 21 MHz Schottky system was originally designed with movable pickup plates to maximize sensitivity [12]. The original incarnation also had two pickups with could be added with a variable phase to suppress the proton tune in favor of the antiproton tune, although the practical usefulness of this feature in operation was very limited. The pickups are resonant with a tunable resonance frequency. This was intended mainly to compensate for the change in capacitance when the plates were moved [13]. In Run II, the plates are left in a fixed position, and hence tuning is only done occasionally.

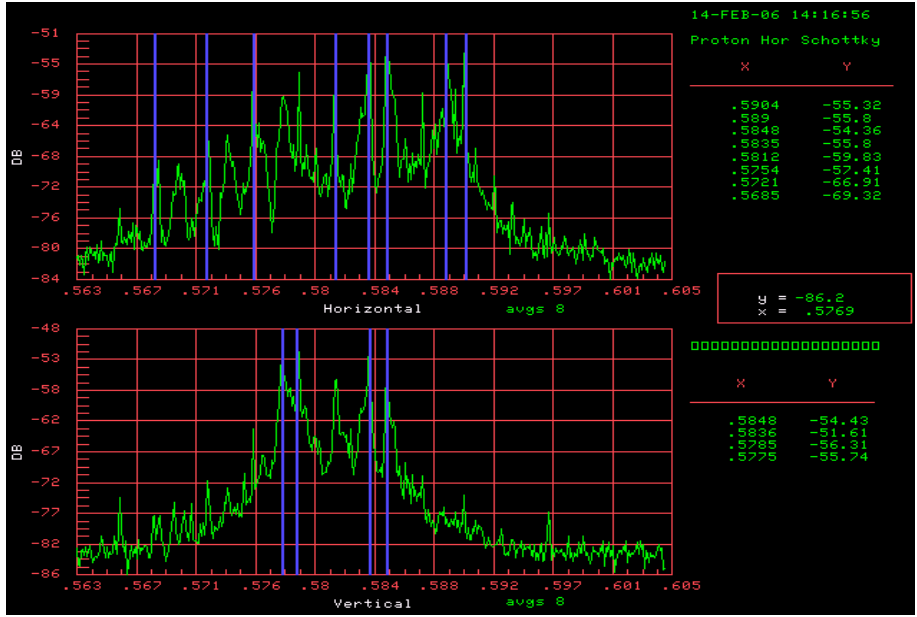

**FIGURE 10.** Schottky spectra from the 21.4 MHz pickups. The spectrum typically contains many peaks, and it may be hard to determine which one is the tune. The system is also unable to resolve antiprotons, due to the much stronger proton signal.

The 1.7 GHz Schottky pickups are slotted waveguide structures (see Fig. 11). The high operating frequency was chosen to be above the coherent spectrum of the beam, thus measuring "true" Schottky signals. Since the devices are not resonant, it is possible to gate on select bunches, making it possible to measure the antiproton tune in the presence of protons. Chromaticity, momentum spread and emittance can also be extracted from the signals, making the 1.7 GHz Schottky a very versatile tool [14].

An advantage of these pickups is that they can be used to measure tunes during normal operation without additional beam excitation. In order to maximize the usefulness of these devices, open access client (OAC) software was developed to run continuously, analyze the data, and publish the resulting tune, chromaticity, momentum spread and emittance on ACNET. Among other things, this allows the tunes to be logged. The 1.7 GHz tune readings are also used in everyday operation to adjust the antiproton tunes as the beam-beam tune shift changes over the course of a store (see Fig. 12).

A peculiarity with the system is that due to the high frequency, the Schottky bands are very wide, and therefore it is not possible resolve the normal modes by frequency. The effect causes an underestimation of the tune separation in the presence of coupling (it can be shown that it approximately measures the uncoupled tunes) [15].

One of the original reasons for developing the system was to be able to extract emittance from the Schottky spectrum during stores. However, it has been observed that even in the microwave range, the Schottky spectrum still have a significant coherent contribution. The reason for this has yet to be fully understood. In the meantime, new thinner carbon filaments have enabled the use of flying wires during stores, reducing the need for Schottky pickups for this particular task [16].

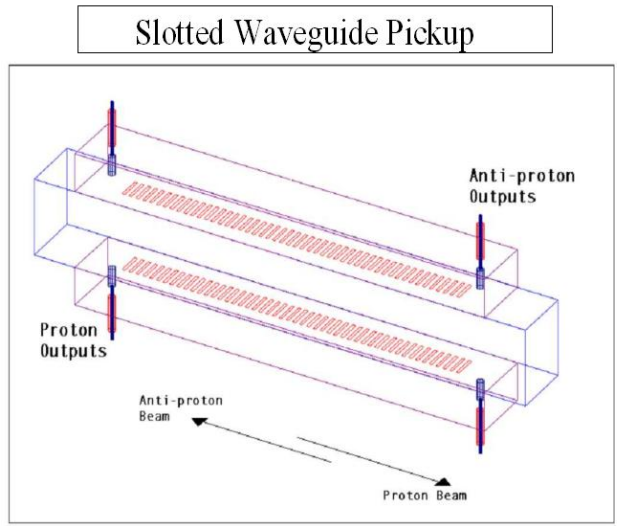

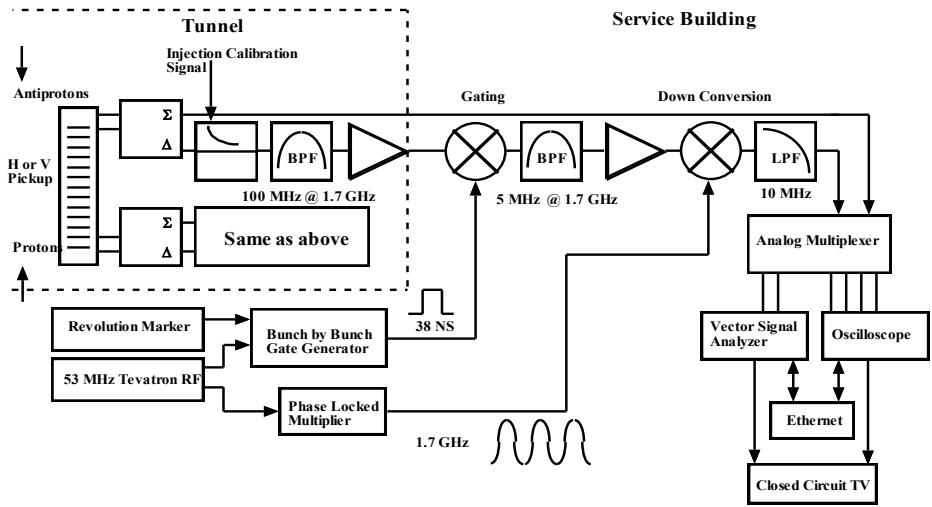

**FIGURE 11.** a) (top) Schematic of a 1.7 GHz slotted waveguide pickup. b) (bottom) Schematic of the electronics and readout system.

It has been estimated that the beam in the Tevatron oscillates with ≈0.1 μm amplitude at betatron frequencies; at lower frequencies, the oscillation amplitudes can be larger ~1-10 μm due to various noise sources, not all well known, including ground motion, jitter in magnet and separator power supplies, and vibrations from the cryogenic system [17]. In an attempt to use this effect to measure the tune without excitation, a very sensitive BPM system has been developed. This system is quite similar to the 3D-BBQ (Direct Diode Detection Base Band Tune) system developed at CERN in that it uses a diode-based sample and hold circuit, but it includes some novel features. Rather than measuring only the positive or negative peak from a stripline doublet, it measures both and takes the difference. It also employs slow feedback to remove baseline variations that can be quite large, thus enhancing the dynamic range.

The system has been successfully tested with proton beam and showed very good tune resolution for individual bunches at a low level of beam excitation by external noise [18].

The original 3D-BBQ system has also been tested using a module provided by CERN. By gating the signal from a stripline, it was able to see antiprotons without beam excitation. However, it suffers from relatively strong 60 Hz lines that have also been observed at RHIC and SPS (in this case 50 Hz, due to the difference in mains frequency) [19].

The tune tracker uses a phase-locked loop (PLL) around the beam response. The beam is excited at a given frequency using a stripline pickup as a kicker, and the response is measured on another stripline. The PLL locks to a given frequency in the tune spectra, defined by a pre-selected phase response value, and tracks any changes in the tune. A novelty in the Tevatron system compared to previous tune tracker implementations is the capability of pulsed excitation. When measured with high resolution, the beam phase response exhibits large excursions from the synchrotron sidebands, which can cause the PLL to jump from one synchrotron band to another. By pulsing the excitation, the measured phase response is smoothed out to follow the slow underlying phase response more closely, resulting in a more reliable measurement [20].

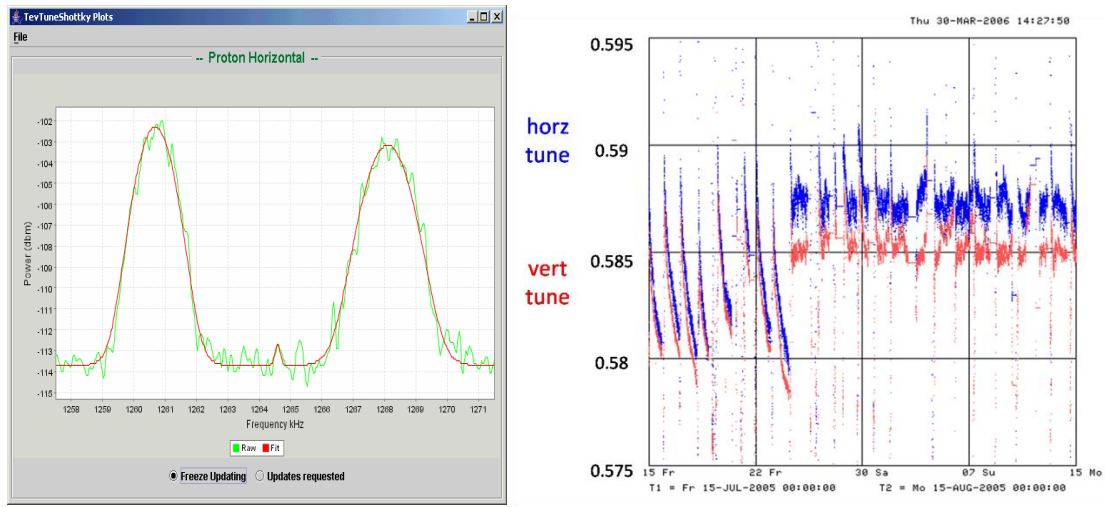

**FIGURE 12.** a) (left) Schottky spectra from the 1.7 GHz Schottky, showing both upper and lower betatron sidebands b) (right) Logged antiproton tune values over several weeks showing the effect of tune compensation during stores. Before the operators started using the 1.7 GHz Schottky tune readback to compensate for tune drifts, the tune would decrease significantly during a store, from ≈0.590 to ≈0.580, as a result of the decreasing beam-beam parameter (top line – horizontal tune, bottom - vertical tune, 0.005 per division).

## Data Logging and On/Offline Presentation

Online and offline access to the vast amount of accelerator data is crucial to evaluating and improving machine performance and diagnosing failures. Retaining bunch-by-bunch values is especially useful since the beam dynamics vary over the bunch positions within a train [1]. In the Tevatron collider complex, the readings and settings of accelerator devices are obtained via Fermilab's own ACNET control system. Device data can be plotted live at up to 1440 Hz. Device data can be logged at various fixed rates or periods, *e.g.* 15 Hz or 1 minute, or on a specific event, *e.g.* when the energy ramp is complete. Logged data is stored in circular buffers on ~70 nodes hosting a MySQL database and ~80 GB of storage for compressed data. The data in the circular buffers wrap-around in a time that depends upon the number of devices and their logged rate for a given logger. Logged data up to a 1 Hz maximum rate is also copied to a "backup" logger for long-term storage.

There are several means of accessing and plotting accelerator data: standard C-based console applications used in operations, Java applications via a web-based interface, exports to Excel spreadsheets and Java Analysis Studio files, as well as programmatic APIs. Each method has its own advantages and disadvantages, but the flexibility allows users, both on-site and off-site, to access the data how they want or need. Fig. 13 shows two examples of accessing Tevatron data.

In addition to the above data logging scheme, data for all Tevatron shots is automatically collected and stored via a package called SDA, for Sequence Data Acquisition [21]. The desired data and plots for all stages of a shot (injection, low-beta squeeze, etc.) can be easily configured. SDA also stands for Shot Data Analysis; SDA software automatically generates summary reports and tables for each store. These data are readily accessible by various means and allow for convenient analysis of the accelerator complex on a shot-by-shot basis [22].

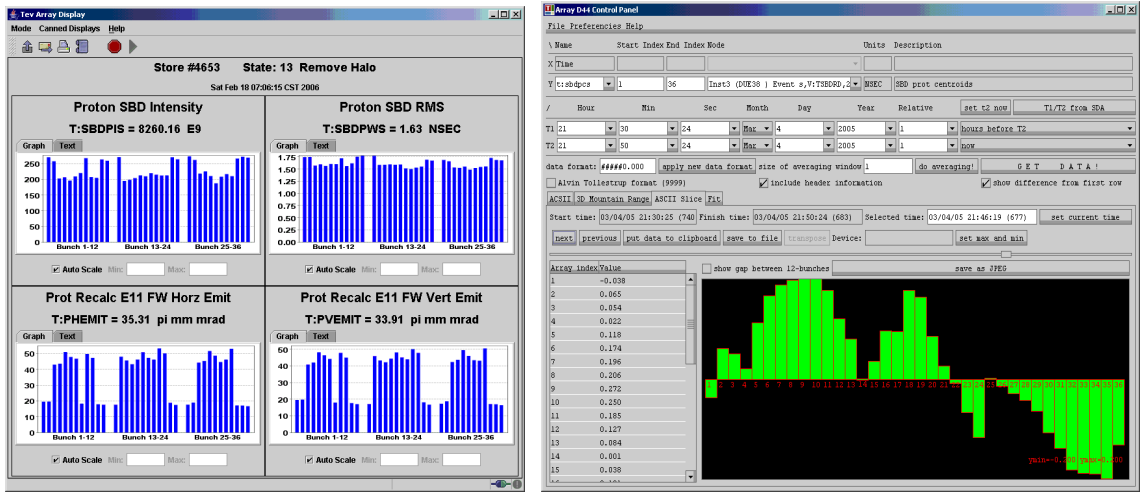

**FIGURE 13.** a) (left) The window of a Java applications showing live, bunch-by-bunch data for Tevatron proton bunches including intensity, RMS bunch length, and transverse emittances; an instability during antiproton injections had caused emittance growth and beam loss for particular proton bunches. b) (right) A Java application showing a snapshot of logged proton bunch centroid positions within their RF buckets; a longitudinal, coupled-bunch instability with ±4 deg of RF phase oscillation amplitude was occurring at the time.

# Longitudinal Beam Diagnostics

The resistive wall monitors used in the Tevatron consists of a short ceramic vacuum pipe with eighty 120 Ω resistors across it. A copper casing around the ceramic break filled with ferrite provides a low impedance bypass for DC currents while forcing AC currents to flow through the resistors. There are also ferrite cores inside the vacuum to improve the signal quality. Signals from four locations around the ceramic pipe are summed to provide an intensity measurement. There are two resistive wall monitors in the Tevatron. One is dedicated to the Fast Bunch Integrator (FBI) and Sampled Bunch Display (SBD) (see Fig. 14), and the other is for general use. This is to avoid errors in intensity readings due to improper terminations.

With the larger number of bunches and higher intensities, the resistive wall monitors developed vacuum problems early in Run II. The cause was beam-related heating of the ferrite cores. This problem was solved by replacing the ferrite with a different type. More recently, the resistive wall monitor used for SBD and FBI developed problems when the surface-mounted resistors over which the signal is measured came loose. This caused step changes in impedance, and therefore calibration. The problem was diagnosed using logged SBD and FBI data, and all of the resistors were replaced.

The longitudinal phase monitor (LPM) [23] is using the signal from a stripline pickup. The original idea was to use the antisymmetric shape of the bunch signal from a stripline pickup, by multiplying it with a sine and cosine function locked to the RF. The two signals are then integrated over the bunch, and the phase can be extracted from the ratio of the two integrals. This was implemented in analog electronics using mixers and gated integrators, and the result was digitized and processed in an FPGA. The FPGA calculated the average over all bunches and output it as an analog voltage through a DAC. Turn-by-turn values were also saved in buffer memory and could be retrieved via an ethernet interface.

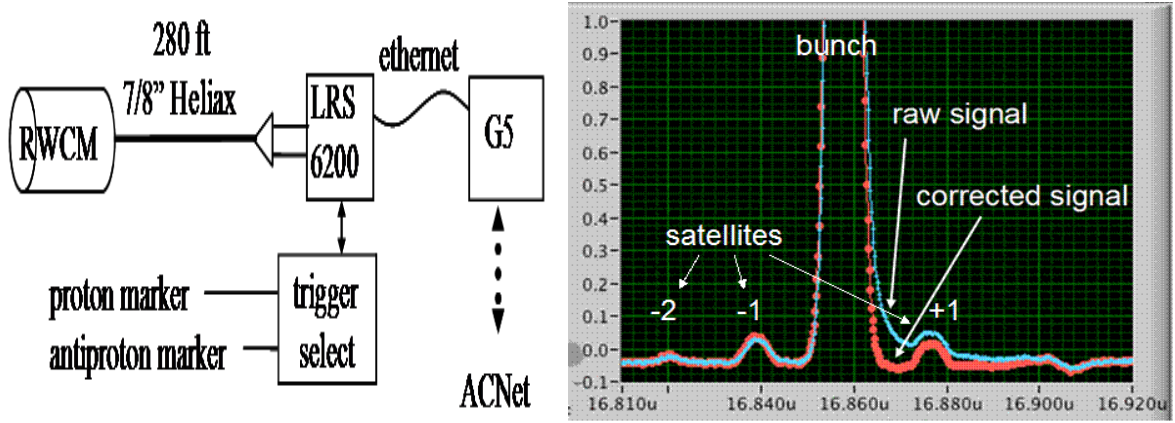

**FIGURE 14.** a) (left) Schematics of the signal acquisition from resistive wall current monitor. b) (right) Raw and corrected proton bunch signal from RWCM.

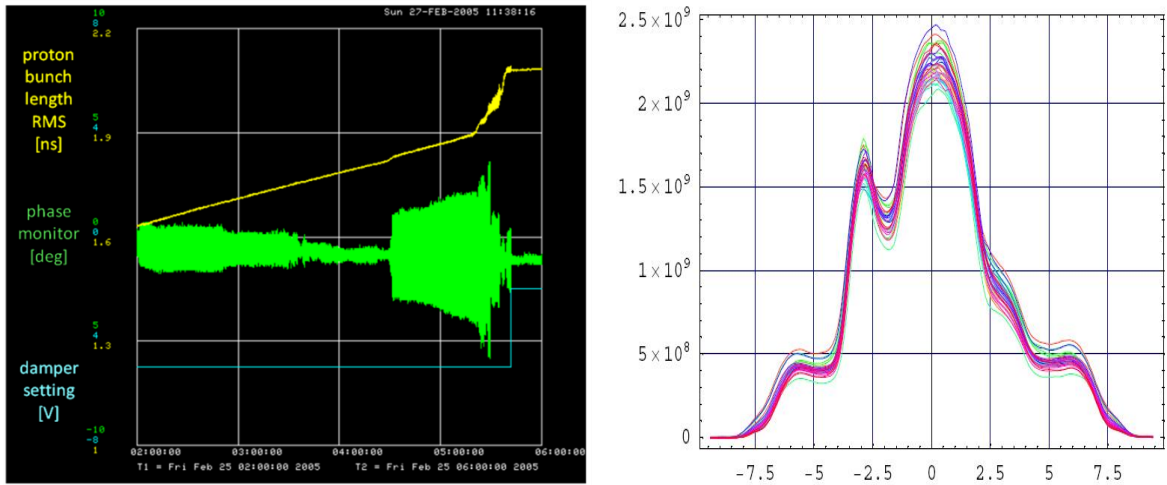

**FIGURE 15.** a) (left) Longitudinal phase monitor readings during an instability. b) (right) Shapes of all 36 proton bunches as detected by SBD after longitudinal instability had developed.

Recently, the longitudinal phase monitor system was redesigned using modified hardware from the bunch-by-bunch baseband tune effort. In this case, the raw signal passed through a 5 MHz Gaussian filter and was then digitized directly. All processing is done digitally. The two integrals are replaced by sums over the part of the signal that arrives before and after a defined "time zero", and the phase extracted from the relative difference between the two [24].

## Abort Gap Monitors

Longitudinal instabilities, RF noise, and intra-beam scattering can cause particles to leak out of RF buckets and into satellites or into the abort gaps [25, 26] – see Fig.15. There are three 2.6 μs gaps between 3 trains of 12 bunches each separated by 396 ns. The presence of even a small fraction (few $10^9$ or 0.0001 of the total) of the beam in the abort gaps can induce quenches of the superconducting magnets, as these particles are sprayed onto the magnets when an abort kicker fires, and inflict severe radiation damage on the silicon detectors of the CDF and D0 experiments. Synchrotron radiation (SR) from these unwanted 980 GeV protons is collected for monitoring their intensity.

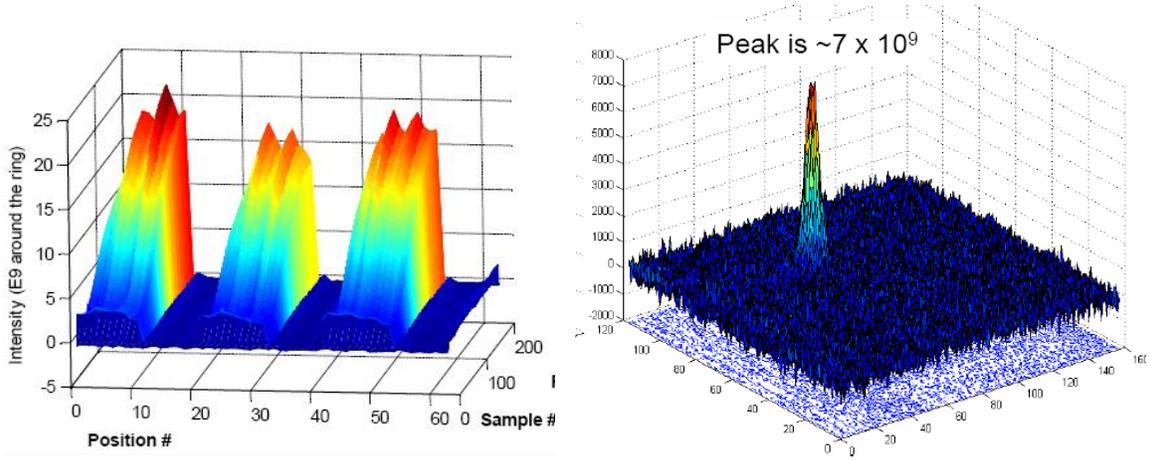

**FIGURE 16.** a) (left) Time evolution of beam in the gaps and between main bunches during a store (sample #200 corresponds to about 20 hrs, position # reflects 340 ns gap synchronization with respect to the revolution marker. Intensity is given in the units of equivalent number of particles if uniformly distributed around the circumference). b) (right) Profile of DC beam in the gap imaged by a CID camera.

A very sensitive gated monitor of the SR from the beam in the gap was developed on the base of Hamamatsu R5916U-50 micro-channel plate (MCP) PMT with a minimum gating time of 5 ns. This tube can be used to measure DC beam intensity immediately following a bunch of protons. The DAQ system consists of a fast integrator, to which the anode of the PMT is connected, and a VME digitizer that is read by an application on a processor board residing in the VME crate. Data is collected for 1000 revolutions and averaged in the processor board. This cycle is repeated every 3 or 4 seconds. The application controls the timing of both the PMT and integration boards. Fig. 16 a) shows how the intensity of the beam outside of the main 36 bunches is growing over a course of a HEP store. A standard synchrotron light monitor equipped with an image intensifier can see the DC beam profile, but only if enough camera frames can be summed together. A LabView-controlled Windows PC system does such integration of CID camera RS-170 video images captured by a frame-grabber card. Fig. 16 b) presents an example of the proton DC beam profile in the Tevatron. Details of calibrating and measuring the intensity of beam in the abort gap using synchrotron light and a gated photomultiplier tube are described in Ref. [27].

## Intensity Measurements

In the Tevatron, a DC Current Transformer (DCCT) and a Resistive Wall Current Monitor (RWCM or RWM) are the pieces of instrumentation that allow beam intensity measurements [28, 29]. The DCCT can only provide a measurement of the total beam intensity (sum of proton and antiproton currents). The RWM does distinguish between protons and antiprotons because of its high bandwidth and location where the protons and antiprotons are well-separated in time.

The DCCT front-end contains an Interactive Circuits and Systems (ICS) ICS-110BL-8B 24-bit, 8-channel ADC to digitize the DCCT signal and a Motorola

MVME-2401 processor. The ADC samples at 6.9 MHz and outputs a 128-sample average measurement at 54 kHz. The crate CPU performs additional averaging and provides the interface to ACNET. There is also a circular buffer that can be stopped upon a beam abort in order to help diagnose the cause of beam loss. The DCCT provides the most precise intensity measurements with a resolution of $\approx 0.5 \times (10)^9$ for typical Tevatron total beam intensities of $10^{11}$ to $10^{13}$ particles. The DCCT is calibrated via an external pulser.

Bunched-beam intensity measurements are made by the FBI and SBD systems, both of which use the RWM as their signal source. The FBI uses ADCs to integrate the RWM output gated on the individual RF buckets and obtain baseline measurements taken in the gaps between each train. A Motorola MVME-2401 processor performs the baseline correction and acts as the interface to ACNET. The FBI system provides narrow-gate (single bucket) and wide-gate (five buckets) intensity measurements for all proton and antiproton bunches at a rate of up to a few hundred Hz. Comparing the narrow and wide-gate values provides a measure of the intensity of satellite bunches, typically a few percent of the main bunch intensity.

The SBD configuration was described previously. The resolution of the bunch intensity measurements is $\approx 0.5 \times (10)^9$ for present typical intensities of 20-80 $\times (10)^9$ for antiprotons and 240-300 $\times (10)^9$ for protons. The SBD can update measurements at approximately 1 Hz rate.

Both the FBI and SBD intensities can be calibrated via the very well-known measurement provided by the DCCT via the equation:

$$I_{DCCT} = I_{P,true} + I_{A,true} = I_{P,RWM} \times (1 + \frac{I_{A,RWM}}{I_{P,RWM}}) \times A_{calib}, \qquad (1)$$

where the DCCT intensity should be equal to the sum of the measured bunch intensities. A few percent correction needs to be made for satellites and other beam observed by the DCCT but not the FBI or SBD. This method requires no knowledge of the RWM, but only the relative gains of the proton and antiproton channels of the system being calibrated.

## Chromaticity Diagnostics

High intensity proton and antiproton beam stability and lifetime depend strongly on machine chromaticity [30]. Fig. 17 demonstrates that proton beam loss is a stronger than linear function of Q'. So, one of the operational challenges in the Tevatron is to measure accurately and control both vertical and horizontal chromaticities so that they are high enough to keep beam stable, yet low enough to avoid high beam losses. Several methods are employed.

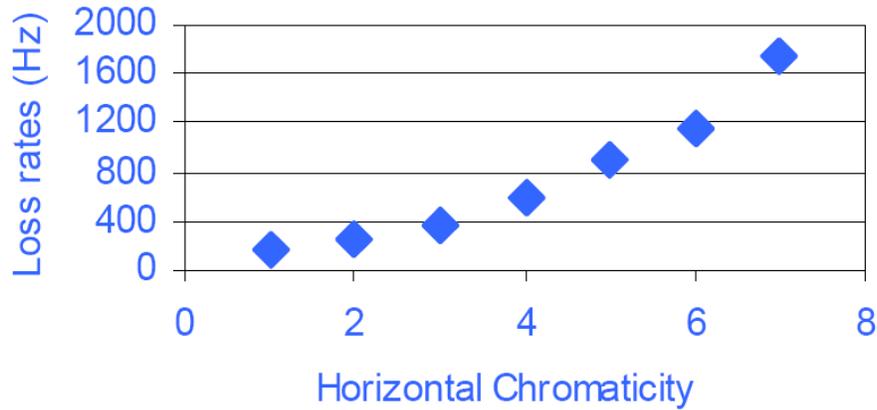

**FIGURE 17.** Loss rate of protons at 150 GeV injection energy versus horizontal chromaticity.

The standard one – observation of the tune change while changing RF frequency – works well and is accurate to ~0.5 unit of Q' with ±40 Hz change of the F_rf=53.1 MHz if measured by 21 MHz Schottky tune detector. A much faster, but just as accurate head-tail method has been developed [31]. In that method, beam is kicked (causing a slight ~5% emittance growth) and the differential motion of bunch head and bunch tail, as measured by a stripline pickup, is recorded by a fast digital scope (Tektronix TDS7000, 1.5 GHz analog bandwidth, 5 GS/s). An example is shown in Fig. 18 a). The amplitude of the motion has a maximum at half of the synchrotron period (about 300 turns in Fig. 18 a), and is proportional to Q'. Another fast technique allowing continuous measurement is based upon RF phase modulation/demodulation using dedicated Chromaticity Tracker (CT) hardware. When modulating the RF phase at a set amplitude and frequency, the chromaticity can be extracted from the relation $Q' = \eta[(k+q_0) - hZ / \Delta\varphi_{mod}]$, where $\Delta\varphi_{mod}$ is the amplitude of the phase modulation, Z is the measured amplitude of the demodulated phase signal, $q_0$ is the measured fractional tune from the Tune Tracker, $\eta = 0.0029$ is the momentum compaction factor, $k = 448$ is the mode number, and $h = 1113$ is the RF harmonic number [32]. Fig.18 b) shows CT measurements of the Tevatron chromaticity drifts at injection energy. The Chromaticity Tracker is now used operationally to set the injection chromaticities for every HEP shot-setup. It has helped make shot-setup faster and reduced variances from shot to shot. The measured chromaticities from the three techniques agree within ~±0.5 unit under the same beam conditions.

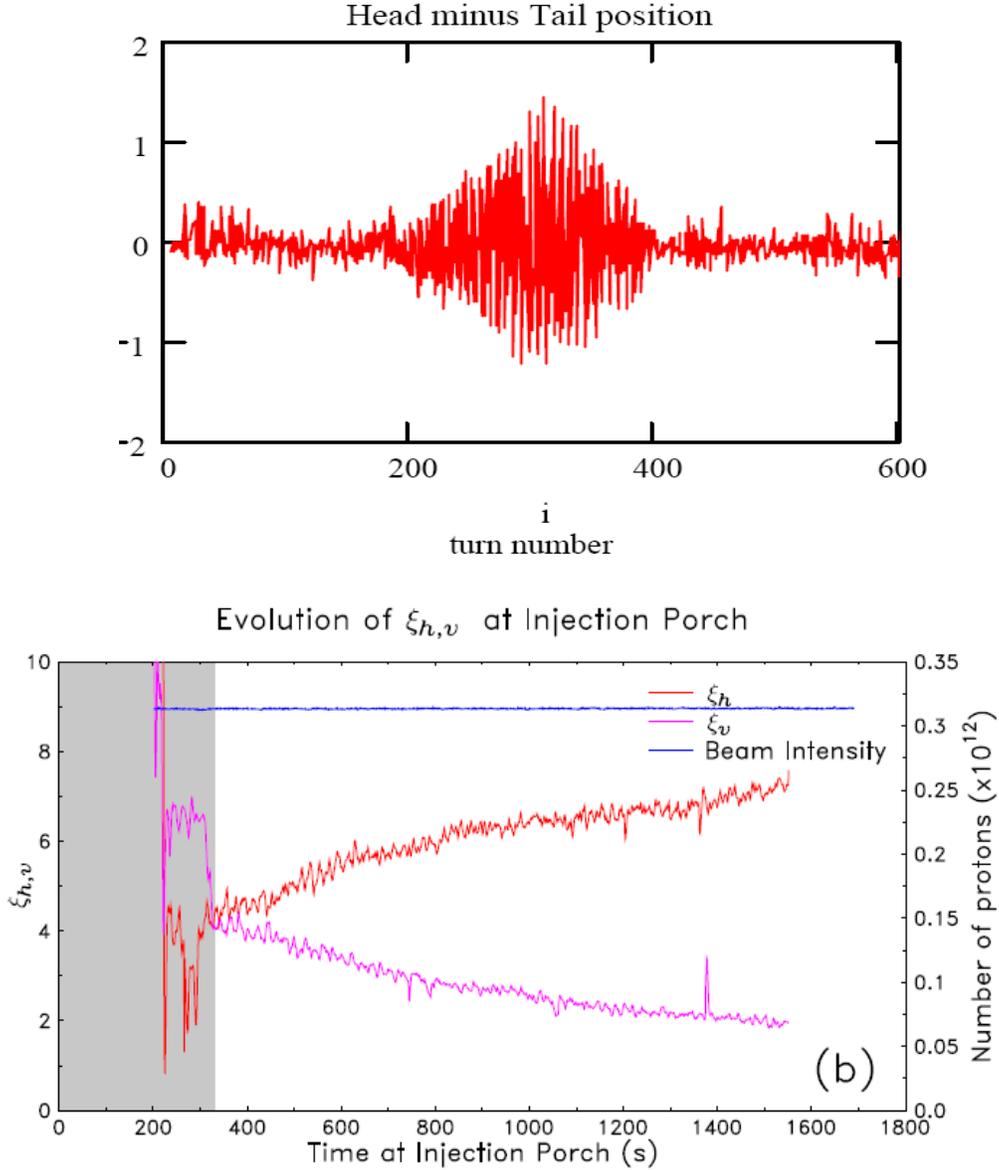

**FIGURE 18.** a) (top) Differential motion (in mm) between the head (+4 ns off bunch center) and tail (-4ns) of a high intensity proton bunch in the Tevatron at 150 GeV vs turn number after a 1 mm vertical kick. b) (bottom) Horizontal and vertical chromaticity drifts in the Tevatron at 150 GeV (with imperfect compensation) as measured continuously by the Chromaticity Tracker. In the gray region, beam was injected and chromaticity set to 4 units in both planes [32].

## Special BPMs and Beam Profile Meters

Flying wires have been the main source of determining transverse emittances and profiles of the protons and antiprotons. There are three flying wire cans in the Tevatron: one horizontal and one vertical at a low dispersion area, and one horizontal at a high dispersion location. The original flying wire cans had 33 μm diameter wires, but those thick wires caused high loss spikes in the experiments when they were used during HEP stores. Thinner, 7 μm wires have been used successfully. The flying wires provide emittance measurements with 1 π mm mrad uncertainty. Uncertainties in the lattice parameters at those locations were a major systematic error for the emittance measurement.

There has been much effort to develop additional means of measuring beam size to verify the flying wire emittances. The synchrotron light monitors [33] have been used, and their performance has been improved through a better understanding of the radiated light, and better imaging hardware and data acquisition.

Two systems that have demonstrated usefulness are the Ionization Profile Monitor (IPM) [34] and the Optical Transition Radiation (OTR) detector [35]. The IPM should allow non-invasive, nearly continuous measurement of the beam profiles at all stages of operation. The OTR was installed close to the IPM and will be used as a cross-check of the IPM at injection; we cannot leave the OTR foils inserted for routine operation. The IPMs are capable of single bunch acquisition for single bunches (see Fig. 19). The OTR could also in principle be used for multi-turn acquisition, but the camera that is currently used does not have enough time resolution. However, by injecting a mis-steered beam, non-overlapping profiles from the two first turns can be obtained (see Fig. 20).

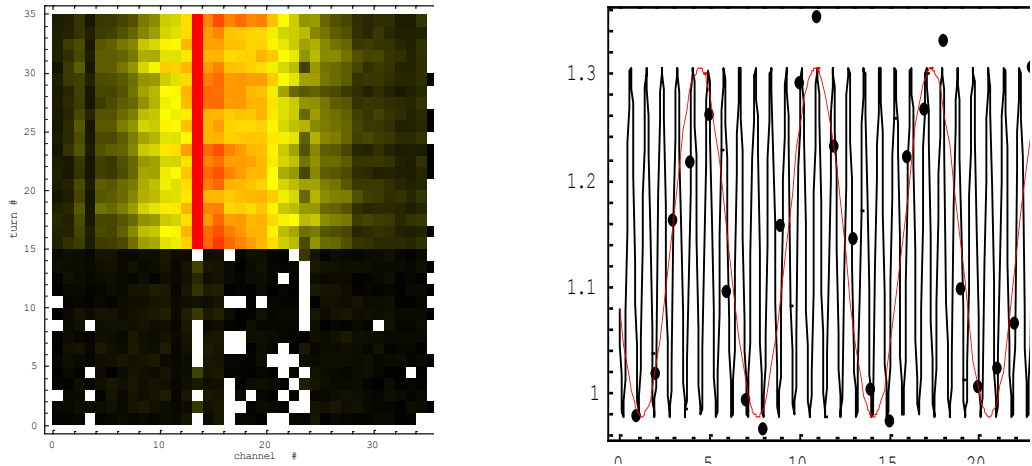

**FIGURE 19.** a) (left) TBT vertical profile of a coalesced proton bunch at injection using the IPM. The bunch was injected at turn #15 on this plot. This horizontal coordinate corresponds to ≈1 cm. b) (right) Measured profile widths calculated from the turn-by-turn data. Black dots are IPM data, red line is a fit to $\cos((2Q_y-41)n)$, black line is a fit to $\cos(2Q_y n)$.

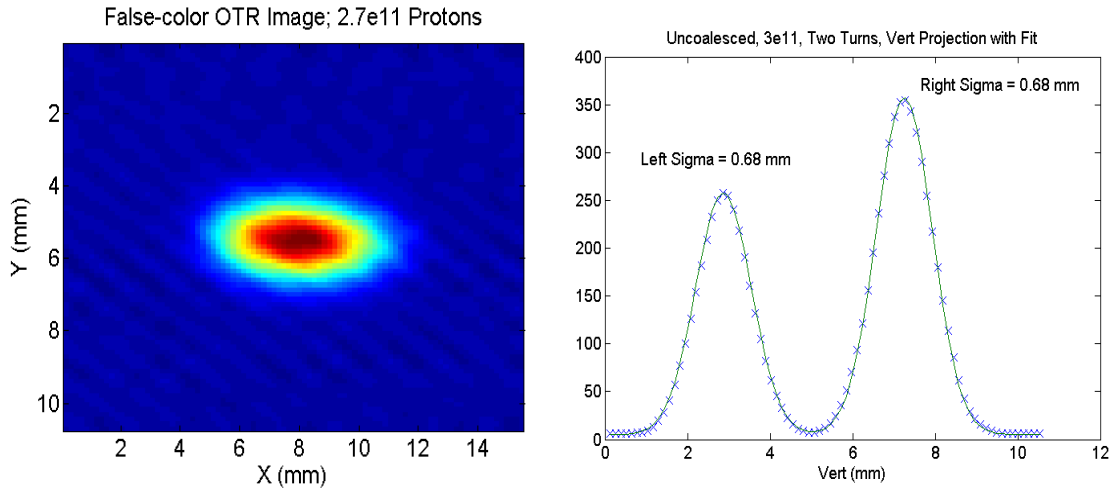

**FIGURE 20.** a) (left) Transverse 2D bunch profile as measured by the OTR  b) (right) Vertical profiles of a single proton bunch from the OTR on two consecutive turns. The second turn profile is offset from the first, and the images are summed together by the slow camera. Note that over the two first turns, the OTR does not show evidence of the quadrupole oscillations seen in the IPM. However, from the IPM data only a 5% effect is expected between these two turns.

To reduce emittance dilution caused by mis-steering at injection, dipole corrector magnets are adjusted by a beam-line tuner system, based on measurements of turn-by-turn orbit positions from directional stripline pickups. The 1 m long striplines are separated by an 83 mm gap, have ≈30 dB directionality and 0.65 dB/mm sensitivity. Measurements from one injection are used to make corrections for the subsequent shot, and usually injection offsets can be reduced from 1 mm to less than ¼ mm. Measurements of synchrotron oscillations can be used to correct energy and RF phase differences between the Main Injector and Tevatron. In addition, the tunes and coupling at injection can also be extracted from the stripline signals. Two different beam-line tuners have been used. In the first, a Tektronix TDS7104 oscilloscope digitizes the sum and difference signals from the striplines, and the embedded PC performs the signal processing. That system has been used only for closure of antiproton injections; it was too slow for the more frequent proton injections. A faster system, based on the Struck SIS3300 digital receiver module, provides 20-40 μm position measurements for 1000 turns [36]. The digitized data is transferred to a PC which performs digital down-conversion at 30 MHz and calculates the positions and time-of-arrival for the transferred bunches. This system can be used for both proton and antiproton injections. During stores, it can also continuously store position data into a circular buffer that is stopped on a beam abort. The buffered data can be used in the post-mortem diagnosis of a lost store.

## CONCLUSION

For the past eight years, Tevatron Collider Run II has been the centerpiece of the HEP program at Fermilab, the US, and the world, and it will continue to be so for the next several years. Such stature led to fixed attention to luminosity progress. Although at the start of the run everything was seemingly in place for successful operation, progress was significantly slower than expected because of unexpected accelerator physics and technology problems all across the board, from antiproton production and beam transfers to dynamics of colliding beams. Mobilization of human and financial resources at Fermilab and assistance from other US DoE laboratories greatly accelerated the resolution of many problems. Development of new diagnostic tools for the Tevatron was needed to provide insights into serious issues of coherent instabilities, beam losses and beam-beam interactions. As a result, almost two dozen various instruments were either developed or significantly improved, and that eventually paid off in the integrated luminosity delivered to the CDF and D0 detectors. At present (Summer 2009), each of the detectors has received nearly 7 $fb^{-1}$ of proton-antiproton collisions at 1.96 TeV center of mass energy, over 40 times more than the luminosity integral for all of Tevatron Run I (1992-1996).

There are several lessons learned during this campaign. First, we realized the importance of multiple instruments for cross-checking and cross-calibrating one another. For example, there are several instruments to measure beam intensity: DC Current Transformer (DCCT), Fast Bunch Integrator (FBI) and Sampled Bunch Display (SBD). The DCCT is the most precise but it has limited application range, *e.g.* it cannot report individual bunch intensities. The FBI and SBD are not as precise but they are really multi-functional, operating on a bunch-by-bunch basis, and calibrating them within 1% of the DCCT made them trustworthy and very useful in operations. In addition, the fast longitudinal phase monitor (LPM) was cross-checked with the SBD. Three tune monitors – 21 MHz Schottky (used for injection tune-up), 1.7 GHz Schottky (most versatile) detectors and Tune Tracker (the fastest and most precise of the three) – are employed in operations for different tasks after being carefully cross-calibrated. A lot of effort over many years was needed to bring the three emittance measurement tools - Flying Wires (FWs), Synchrotron Light Monitor (SyncLite) and 1.7 GHz Schottky detector - into satisfactory agreement; currently they agree within ±5%.

Another lesson is the need for non-invasive beam diagnostics for nearly continuous monitoring of beam parameters. The lack of any natural damping in proton accelerators and the sensitivity of SC magnets to beam losses (quenches) restrict the use of invasive techniques that often have better resolution than non-invasive ones. For example, flying wires is the most precise and understood technique for emittance measurements, but the resulting background spikes and emittance growth limit their use to only once per hour during high-energy collision stores. The complementary, non-invasive Synchrotron light monitor and 1.7 GHz Schottky can report measurements every second.

A third lesson is that the Collider operation team needs fast data collection rate of all diagnostics and control channels (at least 1 Hz) for all channels at all stages of the machine cycle for all bunches all the time – and the data should be saved forever (for years)! That greatly helps to correlate machine behavior now with the past.

We have learned the usefulness of fast access to beam-related information that can be provided by the experimental detectors (CDF and D0, in our case), so good communication between the accelerator and experiment personnel is important. The luminous region parameters information noted above is a good example.

We also have benefited from help and ideas from other groups and laboratories that have expertise in a number of specific areas: for example, Fermilab's Computing Division experts took a leading role in development of DAQ for the Tevatron BPM upgrade; FNAL Particle Physics Division leads Tevatron BLM upgrade and provides luminous region analysis data ($\beta^*$ monitors); Berkeley Lab contributed in the development of the MCP-PMT based Abort Gap Monitor, etc.

And finally, we realized that constructing a new instrument is fast compared to the time needed to make it "fully operational", i.e. satisfactory to operators and physicists. A lot of effort went into the debugging, tune-up, cross-calibration and "polishing" of beam diagnostics. So, we teamed up diagnostics developers and users (physicists and engineers) from the very start of instrument development until the end of its commissioning. Such teams of two to four were very efficient in developing or overhauling about two dozen beam diagnostics instruments for the Tevatron Run II.

We believe that commissioning and operation of the next large colliders – the Large Hadron Collider at CERN and possibly the International Linear Collider and/or a Muon Collider – will set similar demands to beam diagnostics and the lessons we learned at the Tevatron can be taken into account usefully there.

## ACKNOWLEDGMENTS

On behalf of the Run II Collider operations team, we acknowledge numerous contributions to the Tevatron diagnostics development by many people from Fermilab's Accelerator Division Instrumentation Department, RF Department, Controls Department, Operations Department, and Electrical Engineering Support Department. Special thanks to our collaborators from the FNAL Accelerator Physics Center, Computing Division and Particle Physics Division, as well as those from Lawrence Berkeley National Laboratory and from Argonne National Laboratory.